%% file: paper.tex
\documentclass{article}
\usepackage[dvips]{graphicx}
\usepackage{spconf,amssymb,amsmath,epsfig,color}
\newcommand{\beq}{\begin{equation}}
\newcommand{\eeq}{\end{equation}}
\newcommand{\beqn}{\begin{eqnarray}}
\newcommand{\eeqn}{\end{eqnarray}}
\DeclareMathOperator*{\argmin}{arg\,min}
\def\bmath#1{\mbox{\boldmath$#1$}}

\long\def\symbolfootnote[#1]#2{\begingroup%
\def\thefootnote{\fnsymbol{footnote}}\footnote[#1]{#2}\endgroup}

\title{Radio Interferometric Calibration Using a Riemannian Manifold}
\name{Sarod Yatawatta}
\address{ASTRON, The Netherlands Institute for Radio Astronomy,\\ The Netherlands.\\ Email: yatawatta@astron.nl}

\begin{document}
\ninept
\maketitle
\symbolfootnote[0]{$^1$To appear in IEEE ICASSP 2013.}
\begin{abstract}
In order to cope with the increased data volumes generated by modern radio interferometers such as LOFAR (Low Frequency Array) or SKA (Square Kilometre Array), fast and efficient calibration algorithms are essential. Traditional radio interferometric calibration is performed using nonlinear optimization techniques such as the Levenberg-Marquardt algorithm in Euclidean space. In this paper, we reformulate radio interferometric calibration as a nonlinear optimization problem on a Riemannian manifold. The reformulated calibration problem is solved using the Riemannian trust-region method. We show that calibration on a Riemannian manifold has faster convergence with reduced computational cost compared to conventional calibration in Euclidean space.
\end{abstract}
\section{Introduction}
Radio interferometric calibration is the estimation of errors introduced by the propagation medium (such as the ionosphere) and by the receivers (such as the beam shape). In order to produce high fidelity and high dynamic range images, calibration is essential. While contemporary radio interferometric arrays at most have a few tens of receivers (or stations), there is a trend towards building large radio interferometers with hundreds of receivers, an example being the Square Kilometre Array (SKA). This naturally leads to data volumes that are by far greater than what is produced by contemporary radio telescopes.

The maximum likelihood estimation of calibration parameters is in fact a nonlinear optimization problem. Currently, nonlinear optimization algorithms such as the Levenberg-Marquardt (LM) method \cite{Lev44,Mar63} are used in radio interferometric calibration \cite{SAGECAL}. The cost function that is minimized during calibration is invariant to multiplication of the parameters by a 2 by 2 unitary matrix. Therefore, the solutions acquired by calibration will have a unitary matrix ambiguity \cite{H4}.

In this paper, we present the 'quotient manifold' geometry \cite{AMS} of the calibration parameters, which is a better representation of their invariance to multiplication by 2 by 2 unitary matrices. We further develop the geometric structure of calibration parameters, first presented in \cite{interpolation}. Rather than minimizing the cost function in Euclidean space, as is currently done, we minimize the cost function on the developed quotient manifold. We use the Riemannian Trust-Region (RTR) method \cite{RTR} for minimizing the cost function.

Optimization on matrix manifolds has developed significantly during the past decade and a complete overview can be found in \cite{AMS}. In particular, when there is an underlying symmetry in the parameter space (such as the invariance to multiplication by a unitary matrix), exploiting the geometric structure yields better performing algorithms \cite{Mishra,Oja,Mishra1}.

Moreover, algorithms such as the LM operate in real parameter space and the cost of calibration of an interferometric array with hundreds of elements is significant, mainly due to the increased size of the Jacobian \cite{Kaz3}. In this paper, we treat calibration parameters as complex numbers and because we employ the RTR method \cite{RTR}, the computational and memory costs are reduced.
The novelty of the work presented in this paper (relation to prior work) is as follows: (i) We present the quotient manifold geometry of radio interferometric calibration, improving on \cite{interpolation}. (ii) We reformulate radio interferometric calibration as an optimization problem on a Riemannian manifold, where we derive expressions for the Riemannian gradient and the Hessian, following \cite{Mishra}. (iii) We apply the RTR method \cite{RTR} for calibration instead of the traditional Euclidean space calibration algorithms.

The rest of the paper is organized as follows: In section \ref{sec:calib} we give an overview of radio interferometric calibration. Next, in section \ref{sec:geom}, we present the geometric structure of calibration parameters. We present the Riemannian gradient and Hessian operators in section \ref{sec:Rcal} for the calibration cost function. Simulation results are presented in section \ref{sec:results} where we apply the RTR method for calibration and finally, we draw our conclusions in section \ref{sec:conclusions}.

Notation: Matrices and vectors are denoted by bold upper and lower case letters as ${\bf J}$ and ${\bf v}$, respectively. The transpose and the  Hermitian transpose are given by $(.)^T$ and $(.)^H$, respectively. The matrix  Frobenius norm is given by $\|.\|$. The set of real and complex numbers are denoted by  ${\mathbb R}$ and ${\mathbb C}$, respectively. The identity matrix is given by $\bf I$. The matrix trace operator is given by $\rm{trace}(.)$. 
\section{Radio Interferometric Calibration}\label{sec:calib}
In this section, we present radio interferometric calibration as an optimization problem. Consider a radio interferometric array with $N$ receivers. The observed data at a baseline formed by two receivers, $p$ and $q$ is given by
\cite{HBS}
\beq \label{ME}
{\bf V}_{pq}={\bf J}_p {\bf C}_{pq} {\bf J}_q^H + {\bf N}_{pq}
\eeq
where ${\bf V}_{pq}$ ($\in \mathbb{C}^{2\times 2}$) is the observed {\em visibility} matrix. The errors that need to be calibrated for station $p$ and $q$ are given by the Jones matrices ${\bf J}_p,{\bf J}_q$ ($\in \mathbb{C}^{2\times 2}$), respectively. The sky signal (or {\em coherency}) is given by ${\bf C}_{pq}$ ($\in \mathbb{C}^{2\times 2}$). The noise matrix ${\bf N}_{pq}$ ($\in \mathbb{C}^{2\times 2}$) is assumed to have complex, zero mean, circular Gaussian elements.

For an array with $N$ receivers, we can form at most $N(N-1)/2$ baselines that collect visibilities as in (\ref{ME}). We rewrite (\ref{ME}) as
\beq \label{visJ}
{\bf V}_{pq} = {\bf A}_p {\bf J} {\bf C}_{pq} {\bf J}^H {\bf A}_q^T + {\bf N}_{pq}
\eeq
where ${\bf J}$ ($\in \mathbb{C}^{2N\times 2}$) is the augmented matrix of Jones matrices of all stations,
\beq
{\bf J}\buildrel\triangle\over=[{\bf J}_1^T,{\bf J}_2^T,\ldots,{\bf J}_N^T]^T
\eeq
and ${\bf A}_p$ ($\in \mathbb{R}^{2\times 2N}$) (and ${\bf A}_q$ likewise) is the canonical selection matrix
\beq \label{Ap}
{\bf A}_p \buildrel\triangle\over=[{\bf 0},{\bf 0},\ldots,{\bf I},\ldots,{\bf 0}].
\eeq
In (\ref{Ap}), all elements of ${\bf A}_p$ are zero except the $p$-th block which is an identity matrix.

Calibration is the estimation of ${\bf J}$ given the visibilities as in (\ref{ME}). Under a Gaussian noise model, the Maximum Likelihood estimate is
\beq \label{calib}
\widehat{\bf J}=\underset{\bf J}{\argmin} f({\bf J})
\eeq
where the nonlinear cost function $f({\bf J})$ is
\beq \label{fcost}
f({\bf J})\buildrel\triangle\over=\sum_{p,q} \| {\bf V}_{pq} - {\bf A}_p {\bf J} {\bf C}_{pq} {\bf J}^H {\bf A}_q^T\|^2.
\eeq

The sky signal almost always has very little polarization and therefore, the coherencies ${\bf C}_{pq}$ in (\ref{ME}) are diagonal matrices. Therefore, for any unitary ${\bf U}$ ($\in \mathbb{C}^{2\times 2}$), we see that $f({\bf J})=f({\bf J U})$. In other words, for any solution ${\bf J}$, a feasible solution for (\ref{calib}) would also be ${\bf JU}$ where ${\bf U}$ is unitary. Currently, a solution for (\ref{calib}) is obtained by well known nonlinear optimization methods such as the Levenberg-Marquardt \cite{Lev44,Mar63} method and an in-depth overview of current calibration approaches can be found in e.g., \cite{SAGECAL}.
\section{Geometric Structure of Calibration}\label{sec:geom}
In this section, we present the manifold geometric structure of the parameters ${\bf J}$ used in radio interferometric calibration. A manifold can be described as a set of entities, together with a set of mappings (or charts) that can locally describe the manifold in Euclidean space. For a more formal introduction to matrix manifolds, the reader is referred to \cite{AMS}. A problem very similar to what we consider in this section (involving real symmetric positive semi-definite matrices) can be found in \cite{Mishra} and we follow the same approach.

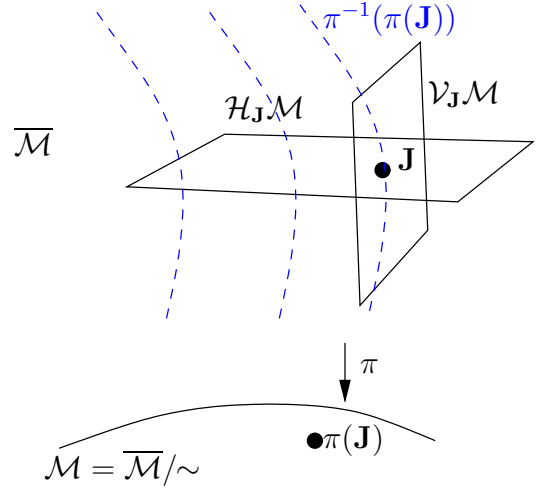
\begin{figure}
\begin{minipage}{0.98\linewidth}
\begin{center}
\input{quotient_manifold.pstex_t}\\
\end{center}
\end{minipage}
\caption{The quotient manifold geometry of the calibration parameters. The dashed (blue) line (on $\overline{\mathcal{M}}$) represents the equivalence class of all solutions that are related to ${\bf J}$ by a unitary ambiguity. This equivalence class is represented by a single point on the quotient manifold ${\mathcal{M}}=\overline{\mathcal{M}}/\!\sim$. The vertical space $\mathcal{V}_{\bf J}\mathcal{M}$ is the vector space tangential to the equivalence class and the horizontal space  $\mathcal{H}_{\bf J}\mathcal{M}$ is the orthogonal complement.\label{quotient_geom}}
\end{figure}

Given the solution to (\ref{calib}), i.e. ${\bf J}$, we know that ${\bf JU}$ is also a feasible solution. We say ${\bf J}$ and ${\bf JU}$ are {\em similar}, i.e.,
\beq \label{sim}
{\bf J} \sim {\bf JU}
\eeq
when ${\bf U}$ is any unitary matrix. Therefore, the whole set of feasible solutions ${\bf JU}$ where ${\bf U}$ is any unitary matrix can be represented by one of its elements, ${\bf J}$. We consider $\overline{{\mathcal{M}}}$ to be the manifold of all $2N\times 2$ complex matrices ($\mathbb{C}^{2N\times 2}$). While the whole set of feasible solutions lie on $\overline{{\mathcal{M}}}$, using the {\em quotient} manifold $\mathcal{M}=\overline{{\mathcal{M}}}/\!\sim$ we can represent the whole set by a single point as shown in Fig. \ref{quotient_geom}.

The mapping $\pi$ (canonical projection) is defined such that any matrix ${\bf JU}$ on $\overline{{\mathcal{M}}}$ is mapped onto a single point, $\pi({\bf J})$ on $\mathcal{M}$. With this mapping, we define the equivalence class
\beq
\pi^{-1}(\pi({\bf J}))\buildrel\triangle\over=\{ {\bf J}{\bf U}: {\bf U}{\bf U}^H={\bf U}^H{\bf U}={\bf I}, {\bf U}\in \mathbb{C}^{2\times 2}\}
\eeq
of solutions represented by a single point on $\mathcal{M}$. In order to make $\mathcal{M}$ a Riemannian manifold, we introduce the (smooth) inner product $g_{\bf J}(.,.)$ to its tangent space ${\mathcal T}_{\bf J}\mathcal{M}$ as
\beq \label{inprod}
g_{\bf J}({\xi}_{\bf J},{\eta}_{\bf J})\buildrel\triangle\over= \mathrm{trace}({\xi}_{\bf J}^H\eta_{\bf J} + \eta_{\bf J}^H \xi_{\bf J}),\ {\xi}_{\bf J},{\eta}_{\bf J} \in {\mathcal T}_{\bf J}\mathcal{M}.
\eeq

With (\ref{inprod}), we can decompose ${\mathcal T}_{\bf J}\mathcal{M}$  into two complementary vector spaces as
\beq
{\mathcal T}_{\bf J}\mathcal{M} = {\mathcal V}_{\bf J}\mathcal{M} \oplus {\mathcal H}_{\bf J}\mathcal{M}
\eeq
where $\oplus$ is the direct sum operator.
We define the vertical space to be the directions tangential to the equivalence class at ${\bf J}$, i.e.,
\beq \label{Vspace}
{\mathcal V}_{\bf J}\mathcal{M}\buildrel\triangle\over=\{ {\bf J}{\bmath \Omega} : {\bmath \Omega}^H=-{\bmath \Omega}, {\bmath \Omega}\in {\mathbb C}^{2\times 2} \}
\eeq
and we choose the set of directions orthogonal to the equivalence class at  ${\bf J}$ as the horizontal space ${\mathcal H}_{\bf J}\mathcal{M}$,
\beq \label{Hspace}
{\mathcal H}_{\bf J}\mathcal{M}\buildrel\triangle\over=\{{\xi}_{\bf J} \in {\mathbb C}^{2N\times 2} : {\xi}_{\bf J}^H {\bf J} = {\bf J}^H {\xi}_{\bf J} \}.
\eeq
The proof of (\ref{Hspace}) is easy to obtain: Let ${\eta}_{\bf J}={\bf J}{\bmath \Omega} \in {\mathcal V}_{\bf J}\mathcal{M}$ then by making $g_{\bf J}({\xi}_{\bf J},{\eta}_{\bf J})=0$, we get (\ref{Hspace}).

The projection of any direction ${\bf Z}\in {\mathbb C}^{2N\times 2}$ onto the horizontal space at ${\bf J}$ is given by
\beq \label{proj}
\Pi_{{\mathcal H}_{\bf J}\mathcal{M}}({\bf Z})\buildrel\triangle\over = {\bf Z}-{\bf J}{\bmath \Omega}
\eeq
where ${\bmath \Omega}$ ($\in {\mathbb C}^{2\times 2}$) is skew-Hermitian and (because ${\bf Z}-{\bf J}{\bmath \Omega} \in {\mathcal H}_{\bf J}\mathcal{M}$) satisfies the Sylvester equation
\beq \label{syl}
{\bmath \Omega} {\bf J}^H{\bf J} + {\bf J}^H{\bf J} {\bmath \Omega} = {\bf J}^H {\bf Z} - {\bf Z}^H{\bf J}.
\eeq

A retraction is a mapping from ${\mathcal T}_{\bf J}\mathcal{M}$ to $\mathcal{M}$. There are many possible retractions but we choose a simple formula for the retraction as
\beq
R_{\bf J}({\xi}_{\bf J}) \buildrel\triangle\over= {\bf J}+{\xi}_{\bf J}.
\eeq

\section{Calibration using a Riemannian manifold}\label{sec:Rcal}
Rather than solving (\ref{calib}) in Euclidean space, we minimize the cost function $f({\bf J})$ on $\mathcal{M}$. In order to do this, we need to compute the Riemannian gradient and the Riemannian Hessian of $f({\bf J})$.
The Riemannian gradient $\mathrm{grad}(f({\bf J}))$ is the unique operator that satisfies
\beq \label{grad}
g_{\bf J}({\xi}_{\bf J},\mathrm{grad}(f({\bf J})))=Df({\bf J})[{\xi}_{\bf J}],\ \forall {\xi}_{\bf J} \in {\mathcal T}_{\bf J}\mathcal{M}
\eeq
where,
\beq
Df({\bf J})[{\xi}_{\bf J}] \buildrel\triangle\over= \underset{t\rightarrow 0}{\mathrm{lim}} \frac{f({\bf J}+t {\xi}_{\bf J})-f({\bf J})}{t}.
\eeq
Using (\ref{fcost}) and (\ref{inprod}), we get
\beqn 
\lefteqn{\mathrm{grad}(f({\bf J}))}\\\nonumber
& = & - \sum_{p,q} \left({\bf A}_p^T({\bf V}_{pq}-{\bf A}_p {\bf J} {\bf C}_{pq} {\bf J}^H {\bf A}_q^T){\bf A}_q{\bf J}{\bf C}_{pq}^H \right.\\\nonumber
&& + \left. {\bf A}_q^T({\bf V}_{pq}-{\bf A}_p{\bf J}{\bf C}_pq{\bf J}^H{\bf A}_q^T)^H {\bf A}_p {\bf J}{\bf C}_{pq}\right)
\eeqn
and the horizontal lift of $\mathrm{grad}(f({\bf J}))$ to ${\mathcal H}_{\bf J}\mathcal{M}$ is 
\beq \label{grad1}
\overline{\mathrm{grad}(f({\bf J}))} =  \Pi_{{\mathcal H}_{\bf J}\mathcal{M}} \left(\mathrm{grad}(f({\bf J}))\right).
\eeq

The Riemannian Hessian is defined as
\beq \label{hess}
\mathrm{Hess}f({\bf J})[\eta_{\bf J}]\buildrel\triangle\over = \Pi_{{\mathcal H}_{\bf J}\mathcal{M}}\left( \underset{t\rightarrow 0}{\mathrm{lim}} \frac{1}{t}\left( \mathrm{grad}f({\bf J}+t\eta_{\bf J})-\mathrm{grad}f({\bf J})\right) \right)
\eeq
where
\beqn
\lefteqn{ \underset{t\rightarrow 0}{\mathrm{lim}} \frac{1}{t}\left( \mathrm{grad}f({\bf J}+t\eta_{\bf J})-\mathrm{grad}f({\bf J})\right)=}\\\nonumber
&&\sum_{p,q}\left( {\bf A}_p^T \left( ({\bf V}_{pq}-{\bf A}_p{\bf J}{\bf C}_{pq}{\bf J}^H{\bf A}_q^T) {\bf A}_q \eta_{\bf J}\right.\right.\\\nonumber
&& \left.\left.- {\bf A}_p({\bf J}{\bf C}_{pq} \eta_{\bf J}^H + \eta_{\bf J}{\bf C}_{pq}{\bf J}^H) {\bf A}_q^T{\bf A}_q{\bf J}\right) {\bf C}_{pq}^H\right. \\\nonumber
&&\left. {\bf A}_q^T \left( ({\bf V}_{pq}-{\bf A}_p{\bf J}{\bf C}_{pq}{\bf J}^H{\bf A}_q^T)^H {\bf A}_p \eta_{\bf J}\right.\right.\\\nonumber
&& \left.\left.- {\bf A}_q({\bf J}{\bf C}_{pq} \eta_{\bf J}^H + \eta_{\bf J}{\bf C}_{pq}{\bf J}^H)^H {\bf A}_p^T{\bf A}_p{\bf J}\right) {\bf C}_{pq}\right). \\\nonumber
\eeqn
Note that for notational purposes we write products such as ${\bf A}_p {\bf J}$ in the above expressions but we do not actually form a matrix product because ${\bf A}_p$-s are merely selection matrices.

With the Riemannian gradient and Hessian at hand, we apply the Riemannian trust-region method \cite{RTR} to our problem. The trust-region method solves the problem
\beq\nonumber
\underset{{\eta}_{\bf J}\in {\mathcal H}_{\bf J}\mathcal{M}}{\mathrm{min}} f({\bf J}) + g_{\bf J}(\overline{\mathrm{grad}(f({\bf J}))},{\eta}_{\bf J})+\frac{1}{2}g_{\bf J}(\mathrm{Hess}f({\bf J})[\eta_{\bf J}],{\eta_{\bf J}})
\eeq
subject to $g_{\bf J}({\eta}_{\bf J},{\eta}_{\bf J})\le \delta^2$, where $\delta$ is the trust-region radius.

The computational cost of the RTR method is significantly less compared with the LM method mainly due to the following reason. In the LM method, with $N$ stations, the Jacobian is a matrix of size $8\times N(N-1)/2$ by $8N$ with real entries. The multiplication of the transpose of the Jacobian with itself has cost $\mathcal{O}\left((8N)^2 4N(N-1)\right)$ and the linear system solved is of size $8N$. On the other hand, in the RTR method, both the gradient and the Hessian are of size $2N \times 2$ with complex entries. Moreover, no full linear system is solved (since the truncated conjugate gradient method is used \cite{RTR,NW}), except in solving (\ref{syl}), which is only a linear system of order $4$.
\section{Simulation Results}\label{sec:results}
In this section we compare the performance of the proposed calibration approach against conventional calibration. For conventional calibration, we consider two optimization algorithms: LM algorithm and Broyden-Fletcher-Goldfarb-Shanno (BFGS) algorithm \cite{NW}. For the LM algorithm, we use closed form Jacobian calculation and for BFGS we use closed form gradient calculation (i.e. not using finite differences). We used the MATLAB implementation of the RTR method \cite{GenRTR} in our simulations.

We simulate an array of $N$ receivers where $N$ is varied. The error matrices ${\bf J}_p,{\bf J}_q$ in (\ref{ME}) are generated with their elements having values drawn from a complex uniform distribution in $[0,1]$ as $\mathcal{U}(0,1)+j\mathcal{U}(0,1)$. The sky signal is kept at unity, i.e.  ${\bf C}_{pq}={\bf I}$. The noise matrix ${\bf N}_{pq}$ is simulated to have complex circular Gaussian random variables. The variance of the noise is changed according to the signal to ratio ($\rm{SNR}$) 
\beq
\mathrm{SNR}\buildrel\triangle\over=\frac{\sum_{p,q} \|{\bf V}_{pq}\|^2}{\sum_{p,q} \| {\bf N}_{pq}\|^2}.
\eeq

The initial values for the parameters are set as ${\bf J}_p={\bf I}$ for $p\in[1,N]$. For the RTR method, the upper bound for the trust region radius $\overline{\delta}$ is chosen as
\beq
\overline{\delta}=\frac{1}{N}\sum_{p,q} \|{\bf V}_{pq}\|^2
\eeq
and the initial trust region radius is chosen as $\overline{\delta}/10$.

\begin{figure}
\begin{minipage}{0.98\linewidth}
\centering
 \centerline{\epsfig{figure=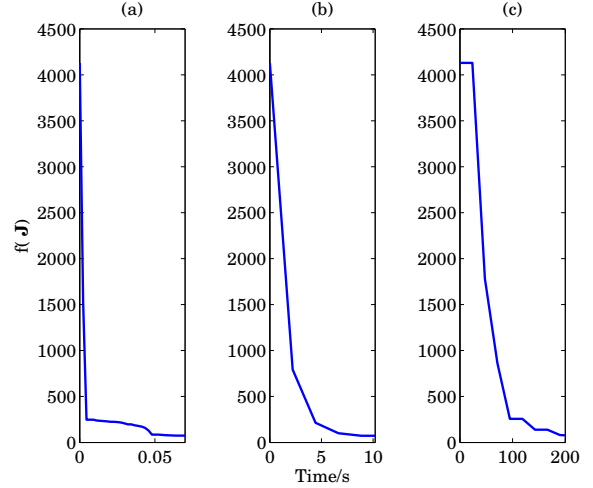,width=9.0cm}}
\caption{Reduction of the cost function $f({\bf J})$ with the time taken for the three optimization algorithms. (a) the RTR method (b) the LM method (c) the BFGS method. The number of stations $N=30$ and the $\rm{SNR}=100$. It is clear that the RTR method uses much less time to reach the minimum cost while both other algorithms take significantly longer times.\label{cost_time}}
\end{minipage}
\end{figure}

In Fig. \ref{cost_time}, we show the reduction of the cost $f({\bf J})$ for $N=30$ and $\rm{SNR}=100$ for the three algorithms. The computing time was measured using a single Intel Xeon CPU core. It is evident that the RTR method takes significantly less time to reach the minimum cost. Furthermore, Fig. \ref{cost_time} shows that both three algorithms reach the minimum cost (i.e. they converge).

\begin{figure}
\begin{minipage}{0.98\linewidth}
\centering
 \centerline{\epsfig{figure=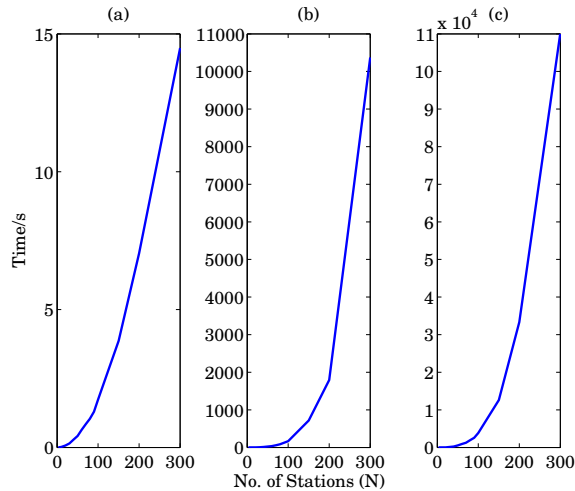,width=9.0cm}}
\caption{Average computation time for the three algorithms for various values of $N$. (a) the RTR method (b) the LM method (c) the BFGS method. The noise is varied with $\rm{SNR}=50, 100, 150,$ and $200$ for each value of $N$. The RTR method takes significantly less time than the other two algorithms.\label{station_time}}
\end{minipage}
\end{figure}

In the next simulation, we vary both $N$ and the $\rm{SNR}$. For each value of $N$, the $\rm{SNR}$ is changed to $50, 100, 150,$ and $200$ and the computation time taken by each algorithm to reach convergence is measured. Once again, we use a single CPU core for the computations. The results are given in Fig. \ref{station_time}. In Fig. \ref{station_time}, we present the average computing time taken for all values of $\rm{SNR}$. The superiority of the RTR method is once again highlighted in this figure. 

\begin{figure}
\begin{minipage}{0.98\linewidth}
\centering
 \centerline{\epsfig{figure=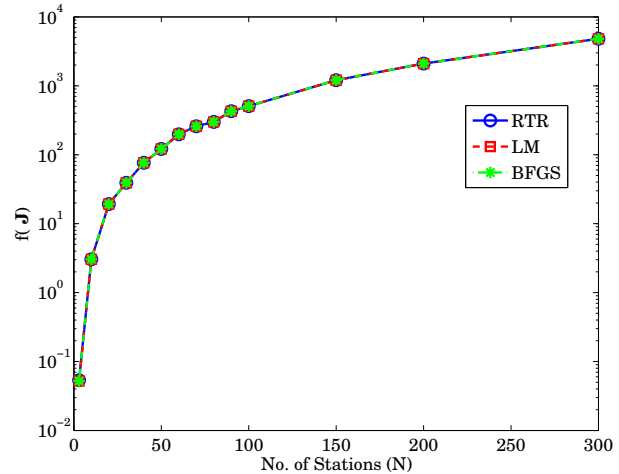,width=9.0cm}}
\caption{Average residual error (or the value of $f({\bf J})$ at convergence) for the three algorithms for various values of $N$. The noise is varied with $\rm{SNR}=50, 100, 150,$ and $200$ for each value of $N$. All three methods attain exactly the same values for $f({\bf J})$ at convergence.\label{station_cost}}
\end{minipage}
\end{figure}

The average residual error for all values of $\rm{SNR}$ (or the value of $f({\bf J})$ at convergence) is shown in Fig. \ref{station_cost}. It is clear that all three methods reach the same final cost at convergence.

\section{Conclusions}\label{sec:conclusions}
We have presented the geometric structure in the form of a Riemannian quotient manifold that can be used in radio interferometric calibration. We have derived the Riemannian gradient and Hessian operators to minimize the cost function used in calibration. By employing the Riemannian trust-region method, we have proposed a computationally efficient calibration method. Based on simulation results, we have shown that the proposed calibration algorithm is much faster and also efficient in memory usage, compared with existing calibration algorithms that operate in Euclidean space.
\bibliographystyle{IEEE}
\bibliography{manref}

\end{document}

%% file: quotient_manifold.pstex_t
\begin{picture}(0,0)%
\includegraphics{quotient_manifold.pstex}%
\end{picture}%
\setlength{\unitlength}{4144sp}%
\begingroup\makeatletter\ifx\SetFigFont\undefined%
\gdef\SetFigFont#1#2#3#4#5{%
  \reset@font\fontsize{#1}{#2pt}%
  \fontfamily{#3}\fontseries{#4}\fontshape{#5}%
  \selectfont}%
\fi\endgroup%
\begin{picture}(3132,2940)(1696,-4090)
\put(1711,-2086){\makebox(0,0)[lb]{\smash{{\SetFigFont{12}{14.4}{\familydefault}{\mddefault}{\updefault}{\color[rgb]{0,0,0}$\overline{\mathcal M}$}%
}}}}
\put(1891,-4021){\makebox(0,0)[lb]{\smash{{\SetFigFont{12}{14.4}{\familydefault}{\mddefault}{\updefault}{\color[rgb]{0,0,0}${\mathcal M}=\overline{\mathcal M}/\!\!\sim$}%
}}}}
\put(3556,-3841){\makebox(0,0)[lb]{\smash{{\SetFigFont{12}{14.4}{\familydefault}{\mddefault}{\updefault}{\color[rgb]{0,0,0}$\pi({\bf J})$}%
}}}}
\put(3556,-1321){\makebox(0,0)[lb]{\smash{{\SetFigFont{12}{14.4}{\familydefault}{\mddefault}{\updefault}{\color[rgb]{0,0,1}$\pi^{-1}(\pi({\bf J}))$}%
}}}}
\put(4006,-2176){\makebox(0,0)[lb]{\smash{{\SetFigFont{12}{14.4}{\familydefault}{\mddefault}{\updefault}{\color[rgb]{0,0,0}${\bf J}$}%
}}}}
\put(3781,-3391){\makebox(0,0)[lb]{\smash{{\SetFigFont{12}{14.4}{\familydefault}{\mddefault}{\updefault}{\color[rgb]{0,0,0}$\pi$}%
}}}}
\put(2971,-1861){\makebox(0,0)[lb]{\smash{{\SetFigFont{12}{14.4}{\familydefault}{\mddefault}{\updefault}{\color[rgb]{0,0,0}${\mathcal H}_{\bf J}{\mathcal M}$}%
}}}}
\put(4186,-1771){\makebox(0,0)[lb]{\smash{{\SetFigFont{12}{14.4}{\familydefault}{\mddefault}{\updefault}{\color[rgb]{0,0,0}${\mathcal V}_{\bf J}{\mathcal M}$}%
}}}}
\end{picture}%